\documentstyle[preprint,revtex,eqsecnum]{aps}
\begin{document}
\draft
\preprint{ }
\begin{title}
Phase separation in the large-spin {\it t}-{\it J} model
\end{title}
\author{ Antimo Angelucci and Sandro Sorella }
\begin{instit}
International School for Advanced Studies, via Beirut 4, 34014 Trieste, Italy
\end{instit}
\begin{abstract}
We investigate the phase diagram of the two dimensional {\it t}-{\it J}
model using a recently developed technique that allows to solve the mean-field
model hamiltonian with a variational calculation. The accuracy of
our estimate is controlled by means of a small parameter $1/q$, analogous
to the inverse spin magnitude $1/s$ employed in studying quantum spin systems.
The mathematical aspects of the method and its connection with other
large-spin approaches are discussed in details.
In the large-$q$ limit the problem of strongly correlated electron systems
turns in the minimization of a total energy functional.
We have performed numerically this optimization problem on a finite but large
$L\times L$ lattice. For a single hole the static small-polaron solution is
stable unless for small values of $J$, where polarons of increasing sizes
have lower energy. At finite doping we recover phase
separation above a critical $J$ and for any electron density, showing that
the Emery {\it et al.} picture represents the semiclassical behaviour
of the {\it t}-{\it J} model. Quantum fluctuations are expected to be very
important especially in the small $J$ -- small doping region, where phase
separation may also be suppressed.
\end{abstract}
\pacs{PACS numbers: 75.10.Jm, 74.20.-z \hfil
      To be published in: the {\it Physical Review B}
}

\narrowtext

\section{ INTRODUCTION }
\label{sec:intro}

The discovery of high-temperature superconductors renewed considerable
interest and attention to the study of strongly interacting electron
systems in one and two dimensions, because it is widely believed that
the anomalous properties of such materials may be related to the strong
Coulomb repulsion and to the low effective dimensionality.

One of the most interesting models which exactly incorporates the constraint
of strong Coulomb repulsion is the well known {\it t}-{\it J} model, which
will be the main topic of the present paper.
Several approximate techniques have been proposed so far to deal with this
model hamiltonian. We mention, for example, the self consistent Born
approximation \cite{klr}, the semiclassical approach \cite{ss},
and recently the limit of infinite dimensionality \cite{voll},
which have been developed
for the one-hole case. In the general case of arbitrary density, the large-$N$
expansion with slave-boson and the slave-fermion techniques \cite{grilli}
should also be mentioned, as well as the vast amount of numerical and
variational works \cite{numw}.
The results of all these different methods are quite controversial, especially
concerning the question about the presence in the model of superconductivity
and/or of marginal Fermi liquid behaviour.

Quite recently Emery {\it et al.} \cite{emery} tried to explain the full phase
diagram of the model, basing their analysis both on a variational argument in
the small $J$ region and on some numerical evidences emerging from a
$4 \times 4$ lattice exact diagonalization.
They speculated that {\sl phase separation} should occur throughout the full
phase diagram: the electron-system phase separates into a phase where the
holes move in a fully polarized state and into an electron-rich phase
characterized by a Ne\`el antiferromagnetic spin-order.
Accordingly, phase separation occurs below some critical doping $\delta_c$,
where $\delta_c \to 0$ for $J \to 0$. Castellani {\it et al.} \cite{castro}
then argued that superconductivity may occur close to the phase separation
boundary. A numerical work by Ogata {\it et al.} \cite{ogata}
provided the complete determination of the one dimensional phase diagram and
evidenced this property in one dimension.

After the Emery's work, Putikka {\it et al.} \cite{putikka} performed a
systematic high-temperature expansion on the {\it t}-{\it J} model and found
evidence that the separated phase should appear only for large values of $J$.
Although this work is surely non-conclusive, they pointed out that
there is not convincing evidence that the separated phase should be
continuously connected through the whole phase diagram, because the small
$J$ and the large $J$ region should behave in a very different way.

Due to the present controversy, we derive here a consistent mean-field
calculation on the {\it t}-{\it J} model using a recently developed
technique \cite{al} allowing to control the accuracy of the mean-field
estimate by means of a small parameter $1/q$, which is similar to the
parameter $1/s$ -- the inverse spin magnitude -- used in the spin-wave
theory of quantum spin systems. This technique
has the following advantages: $i$) does conserve the symmetries of the
{\it t}-{\it J} model hamiltonian, $ii$) takes exactly into account the
constraint of no-double occupancy, and $iii$) ensures that
the classical estimate of the energy is
{\sl variational} for any value of $q$. In this way we generalize a very
useful property known to hold for spin systems, i.e., the classical solution
is independent of the spin magnitude and the corresponding energy is
therefore variational.

With our mean-field approach we are able to reproduce the phase separation
over the small and the large $J$ region, supporting at the semiclassical
level the Emery {\it et al.} picture.
We have made an intensive and systematic numerical work in order
to calculate the true phase diagram at the mean-field level without imposing
any {\sl a priori} order parameter. At the end the phase diagram looks
extremely simple, but unfortunately poor.
However our calculation represents  only the first step towards a complete
determination of the phase diagram. As standard
in any semiclassical approach, the second step would be to include quantum
fluctuations about the mean-field. With the present technique corrections
to the classical energy can be introduced systematically.

The paper is organized as follows: in Sec.\ \ref{sec:coh} we review the graded
Holstein-Primakoff map \cite{al} for the graded algebra spl(2,1) and present
the associated coherent states.
These mathematical tools are developed starting from the observation that
spl(2,1) is the algebra of the operators entering the strongly correlated
electron systems and that it has a series of representation characterized
by a parameter $q$ analogous to the spin magnitude.
The small parameter $1/q$ allows to systematically define distinct large-spin
limits of the {\it t}-{\it J} model. In Sec.\ \ref{sec:large-spin}
we investigate the physical
implications underlying the different generalizations of the model and show
that previously proposed effective hamiltonians, notably the
Kane-Lee-Read \cite{klr} and the spinless fermion hopping hamiltonian
\cite{wen}, can be derived as particular cases of our approach. In Sec.\
\ref{sec:mean-field} we then derive the variational total energy functional
which solves our mean-field theory and present the results concerning the
numerical investigation. Sec.\ \ref{sec:discuss} contains the conclusions.

\section{ ${ spl}$(2,1) COHERENT STATES }
\label{sec:coh}

The {\it t}-{\it J} model hamiltonian \cite{tjm} is defined by
\begin{equation}
H_{tJ}=t\sum_{ij,a}\Omega_{ij}{\tilde c}^{\dagger}_{ai}{\tilde c}_{aj}+
   {J\over2}\sum_{ij}\Omega_{ij}\left({\vec S}_i{\vec S}_j-{1\over 4}
   {\hat N}_{i}{\hat N}_{j}\right), \label{tjmod}
\end{equation}
where $a=1,2$ are spin up and spin down indices, respectively; $\Omega_{ij}$
is the adherence matrix with periodic boundary conditions connecting nearest
neighbors sites, $i,j$ denote the lattice coordinates, and
\begin{mathletters}
\begin{eqnarray}
{\tilde c}_{1i} & = & (1-c_{2i}^{\dagger}c_{2i})c_{1i}\; , \;
{\tilde c}^{\dagger}_{1i}=c_{1i}^{\dagger}(1-c_{2i}^{\dagger}c_{2i})\, ,
\nonumber \\
{\tilde c}_{2i} & = & (1-c_{1i}^{\dagger}c_{1i})c_{2i}\; , \;
{\tilde c}^{\dagger}_{2i} =
c_{2i}^{\dagger}(1-c_{1i}^{\dagger}c_{1i})\, , \label{oddop}
\end{eqnarray}
are hole creation/annihilation operators and
\begin{eqnarray}
{\hat N}_i & = & (c_{1i}^{\dagger}c_{1i}+c_{2i}^{\dagger}c_{2i})\; , \;
S_{3i}={1\over2} (c_{1i}^{\dagger}c_{1i}-c_{2i}^{\dagger}c_{2i})\, ,
\nonumber \\
S_{1i} & = & {1\over2} (c_{1i}^{\dagger}c_{2i}+c_{2i}^{\dagger}c_{1i})\; , \;
S_{2i}={1\over2i}(c_{1i}^{\dagger}c_{2i}-c_{2i}^{\dagger}c_{1i})\, ,
\nonumber \\
 &   & \label{evenop}
\end{eqnarray}
\end{mathletters}
are charge and spin operators, where $c_{1i}^{\dagger}$, $c_{2i}^{\dagger}$
are the electron operators, and doubly occupied sites are excluded. The
exchange constant is always positive: $J\ge 0$.

The $3\times 3$ Hubbard matrices are defined by taking the expectation values
of the operators (2.2) in the restricted single-site Hilbert space
$\vert 0_i \rangle$,
$\vert 1_i \rangle = c_{1i}^{\dagger}\vert 0_i \rangle$,
$\vert 2_i \rangle = c_{2i}^{\dagger}\vert 0_i \rangle$
\begin{eqnarray}
\chi_{ai} & =  &
\langle \alpha_i \vert \,{\tilde c}_{ai}\, \vert \beta_i \rangle\; , \;
{C}_{i}     =
\langle\alpha_i\vert \, {\hat N}_i\, \vert\beta_i\rangle\; , \nonumber \\
\chi^a_i   & = &
\langle\alpha_i\vert \,{\tilde c}^{\dagger}_{ai}\,\vert\beta_i\rangle\; , \;
{\vec Q}_i   =
\langle \alpha_i \vert \,{\vec S}_i\, \vert \beta_i \rangle\; , \;
\alpha,\beta=0,1,2\; \nonumber \\
   &  &  \label{hubbmat}
\end{eqnarray}
The matrices $ \chi^a_i $ and $ \chi_{ai} $ are conjugate
-- i.e., $ \chi^a_i = \chi_{ai}^{\dagger} $. However for later
convenience we prefer to use the notation with the raised index. The matrices
(\ref{hubbmat}) are useful because the constraint of no doubly occupancy can
be automatically taken into account by replacing in Eq.\ (\ref{tjmod}) the
operators (2.2) with the corresponding matrices (\ref{hubbmat}).
Defining $Q_{0i}\equiv (I-{1\over2}{C}_{i})$, where $I$ is the
identity matrix, the Hubbard matrices (\ref{hubbmat}) can be divided
into odd (i.e., ``fermionic'') generators, $\chi_{ai}$, $\chi^a_i$, and even
(i.e., ``bosonic'') generators
$Q_{\mu i}=(Q_{0i},{\vec Q}_i)$ -- which we collect in four-vectors --
satisfying the commutation/anticommutation relations of the spl(2,1)
graded algebra \cite{al,wieg}
\begin{mathletters}
\begin{eqnarray}
&& [Q_{\mu i},Q_{\nu j}]  =  \delta_{ij}
i{\epsilon_{0\mu\nu}}^{\lambda}Q_{\lambda j}\; , \label{algebraee} \\
&& [\chi^a_i,Q_{\mu j}]   =  \delta_{ij}
{1\over 2}{(\sigma_{\mu})_b^{\, a}}\chi^b_j\; ,  \label{algebraeo} \\
&& \{\chi_{ai},\chi^b_j\}  =  \delta_{ij}
   {(\sigma^{\mu})_a^{\, b}}Q_{\mu j}\; ,  \;
  \{\chi_{ai},\chi_{bj}\}  =   0\; ,  \label{algebraoo}
\end{eqnarray}
\end{mathletters}
where the completely antisymmetric tensor is normalized by
${\epsilon_{012}}^{3}=1$; whereas the spin four-vector is built from
the standard Pauli matrices as
$(\sigma^{\mu})_a^{\, b}\equiv (\delta_{ab},{\vec \sigma}_{ab})$.
The greek four-vector indices are raised and lowered using the metric tensor
$g_{\mu\nu}={\rm diag}(1,-1,-1,-1)$, and a summation over repeated ``Lorentz''
and latin (``spinorial'') indices is understood. The generator $Q_{0i}$ has to
be introduced in order to close the spl(2,1) algebraic rules
(\ref{algebraeo},\ref{algebraoo}).
For the time being our considerations will refer mostly to single-site
quantities and to simplify the notation the site index will be dropped
-- if not otherwise needed.

As explained in details in Ref.\ \cite{snr}, the several classes of irreducible
representations of the spl(2,1) graded algebra are labelled by the
eigenvalues of the operators $Q_0$, ${\vec Q}^2$, and $Q_3$, respectively
denoted $q_0$, ${q(q+1)}$ ($q$ is called isospin, and is an integer or
half-integer number), and $q_3$.
The basis vectors are thus denoted with $\vert q_0,q,q_3\rangle$.
Among them, those relevant for our investigation are the so called atypical
representations.
They are characterized by a linear relation between $q_0$ and $q$ and are thus
identified only by the value of the isospin $q$ -- this being the parameter
which we shall use analogously to the spin magnitude $s$ in the standard
spin-wave theory of the Heisenberg hamiltonian.
In each such atypical representation, of dimensionality $(4q+1)$, the basis
vectors can be grouped into two multiplets of $(2q+1)$ and $2q$ states
with quantum numbers
\begin{mathletters}
\begin{eqnarray}
&& \vert q,q,q_3\rangle, \;\; q_3=-q,-q+1,\dots,q, \label{veceven} \\
&& \vert q+{1\over2},q-{1\over2},q_3\rangle, \;\;
q_3=-q+{1\over2},\dots,q-{1\over2}; \label{vecodd}
\end{eqnarray}
\end{mathletters}
which we assume as even and odd, respectively (the grading of the states is
a pure matter of convention). The relative normalization of the two multiplets
is not {\sl a priori} fixed. For the isospin value $q={1\over 2}$ the states
(2.5) are the real spin-${1\over 2}$ and hole (spin-0) states, respectively,
and the matrices defining the representation turn in the $3\times 3$ Hubbard
matrices. Thus the operators (2.2) entering
the {\it t}-{\it J} model belong to the fundamental $q={1\over 2}$ atypical
representation of spl(2,1). According to Eqs.\ (2.5), models of strongly
correlated electron systems can be generalized in a natural way by enlarging
the dimensionality of the spl(2,1) representation. In particular we are
naturally lead to identify the multiplet (\ref{veceven}) as the generalized
spin state, and the multiplet (\ref{vecodd}) as the generalized hole state.
The different value of $q_0$ between the two multiplets is related to the
grading. Defining the ``fermion number operator'' $F=2Q_{0}-2qI$ -- where
$I$ is the unit matrix of the $q$-representation -- then the spin multiplet
(\ref{veceven}) has $ F = 0 $ and the hole multiplet (\ref{vecodd}) has
$ F = 1 $.
Thus for a many-particle system the conservation of the number of particles can
be rewritten as the conservation of the
fermion number
\begin{equation}
     \sum_{i=1}^N F_i =
     \sum_{i=1}^N (2Q_{0i}-2qI) = N_h= N-N_{\rm el} ;
\label{conserv}
\end{equation}
where $N_{\rm el}$ is the number of electrons, $N$ is the number of sites,
and thus $N_h$ is the number of holes.

Assuming the maximal isospin state as the reference vacuum,
$\vert 0\rangle\equiv\vert q,q,q\rangle$ (not to be confused with the
electronic empty state), and introducing a canonical boson $a$ and a canonical
spinless fermion $c$, then the $(4q+1)$-dimensional Hilbert space spanned by
the states (2.5) can be put into correspondence with
the Fock space generated by the states
$\vert B,n\rangle=(a^{\dagger})^n\vert 0\rangle$ [corresponding to
Eq.\ (\ref{veceven})],
and $\vert F,n\rangle=c^{\dagger}(a^{\dagger})^n\vert 0\rangle$ [corresponding
to Eq.\ (\ref{vecodd}), with
$\vert F,0\rangle = \vert q+{1\over2},q-{1\over2},q-{1\over2}\rangle $],
satisfying the operatorial relation $a^{\dagger}a+c^{\dagger}c\le 2q$.

In this basis, the generators (2.4) can be represented as follows \cite{al}
\begin{eqnarray}
&& Q_0=q+{1\over 2}c^{\dagger}c\; , \; \;
   \chi_1=c^{\dagger}\sqrt{2q-a^{\dagger}a-c^{\dagger}c} \; , \nonumber \\
&& Q_3=q-a^{\dagger}a-{1\over 2}c^{\dagger}c\; , \; \;
   \chi^1=\sqrt{2q-a^{\dagger}a-c^{\dagger}c}~c\; , \nonumber \\
&& Q_+=\sqrt{2q-a^{\dagger}a-c^{\dagger}c}~a\; , \; \; \;
   \chi_2=c^{\dagger}a\; , \nonumber \\
&& Q_-=a^{\dagger}\sqrt{2q-a^{\dagger}a-c^{\dagger}c}\; , \; \; \;
   \chi^2=a^{\dagger}c\; , \label{hpmap}
\end{eqnarray}
where $Q_{\pm}=Q_1\pm i Q_2$. Note that the representation of $\chi_1$
(and $\chi^1$) can be simplified -- i.e., we have
$\chi_1=c^{\dagger}\sqrt{2q-a^{\dagger}a}$. The symmetric notation
(\ref{hpmap}) is useful to check that Eq.\ (\ref{hpmap}) satisfy to Eqs.\
(2.4). These equations represent the generalization to spl(2,1) of the usual
Holstein-Primakoff (HP) transformation \cite{hp} for su(2) and in this respect
are similar to those of the Swinger bosons/slave-fermion
representation \cite{aul}. Using the realization (\ref{hpmap}) of the
generator $Q_0$, the conservation law (\ref{conserv}) becomes the conservation
of the spinless fermion number operator:
${\hat N}_h \equiv  \sum_{i}c_i^{\dagger}c_i=N_h$,
so that at half-filling (i.e., $N_h=0$) Eq.\ (\ref{hpmap}) reduce to the
standard su(2) HP transformation
\begin{eqnarray}
  Q_3 & = & s-a^{\dagger}a\, , \nonumber \\
  Q_+ & = & \sqrt{2s-a^{\dagger}a}~a\, \,      \label{hpstandard} \\
  Q_- & = & a^{\dagger}\sqrt{2s-a^{\dagger}a}\, ,  \nonumber
\end{eqnarray}
where $a^{\dagger}a\le 2s$ (using the standard symbol $s$ instead of $q$),
in agreement with the result that at half-filling the
{\it t}-{\it J} model becomes the antiferromagnetic (AF) Heisenberg model.

As for spin systems, we wish to describe spin configurations classically.
To this aim we need the transformations
$(Q_{\mu},\chi_a)\rightarrow (P_{\mu},X_a)$ of the generators which
preserve the spl(2,1) algebra. This can be conveniently done by choosing
the original fermionic representation (2.2) of the $q={1\over 2}$ generators
(2.3), and then considering the most general transformation of the electron
operators which preserves the canonical commutation relations (2.4) as well
as the constraint of no double occupancy. Simple algebra shows that the odd
generators transform like su(2) operators in the fundamental representation:
\FL
\begin{eqnarray}
&& X_1 = e^{i\gamma}e^{-i{\phi\over2}}\left(
  \cos{{\theta\over2}} e^{ i{\psi\over2}}\chi_1 -
  \sin{{\theta\over2}} e^{-i{\psi\over2}}\chi_2\right)\, , \nonumber \\
&& X_2 = e^{i\gamma}e^{i{\phi\over2}}\left(
  \sin{{\theta\over2}} e^{ i{\psi\over2}}\chi_1 +
  \cos{{\theta\over2}} e^{-i{\psi\over2}}\chi_2 \right)\, , \nonumber \\
&& X^1 = X_1^{\dagger}\, , \; \; X^2 = X_2^{\dagger}\, , \label{oddtransf}
\end{eqnarray}
whereas the even generators transform as
\begin{equation}
     P_0 = Q_0\; , \; P_k = R_{km} Q_m ,  \label{eventransf}
\end{equation}
where $R_{km}$ is the SO(3) rotation matrix, whose explicit realization can
be easily derived by means of the anticommutation rule
(\ref{algebraoo}). By retaining
only the leading order terms of the HP transformation (\ref{hpmap}) we
have $Q_0=Q_3 \to q$, $Q_{\pm} \to 0$, $\chi_1 \to {\sqrt{2q}}\,c^{\dagger}$,
$\chi_2 \to 0$, so that in the large-$q$ limit the rotated even
generators (\ref{eventransf}) become classical $c$-numbers
\begin{eqnarray}
&& P_{\mu} \to q(1,{\vec n})\equiv p_{\mu}\; , \nonumber \\
{\vec n} = && (\sin\theta\cos\phi,\sin\theta\sin\phi,\cos\theta),
\label{classvec}
\end{eqnarray}
and the rotated odd generators (\ref{oddtransf}) become proportional to
the spinless fermion operator
\begin{eqnarray}
&&   X_1 \to \sqrt{2q} \, e^{i\omega}e^{-i{\phi\over2}}
     \cos({\theta\over2}) c^{\dagger}\; , \;\;  X^1=X_1^{\dagger},\nonumber \\
&&   X_2 \to \sqrt{2q} \, e^{i\omega}e^{ i{\phi\over2}}
     \sin({\theta\over2}) c^{\dagger}\; , \;\;  X^2=X_2^{\dagger}\; ,
\label{spinless}
\end{eqnarray}
where we have set $\omega=\gamma+\psi/2$.

In the previous equations $\theta$ and $\phi$ are the physical angles which
parametrize the classical spin configuration [see Eq.\ (\ref{classvec})].
Instead, the third Euler angle $\psi$ and the global fermionic phase $\gamma$
are unessential. In fact, $\psi$ can always be eliminated by redefining
$ {\tilde Q_{\pm}}\equiv e^{{\mp}i\psi}Q_{\pm}$,
${\tilde \chi_2}\equiv e^{-i\psi/2}\chi_2 $,
${\tilde \chi_1}\equiv e^{i\psi/2}\chi_1 $, and $\gamma$ can be arbitrarily
fixed by changing the relative normalization of the two multiplets (2.5).

Similarly to spin models, we can give to
Eqs.\ (\ref{classvec},\ref{spinless})
a quantum mechanical meaning by introducing the even coherent state
$\vert \Omega_b \rangle$ of the generalized spin multiplet (\ref{veceven}) and
the odd coherent state $\vert \Omega_f \rangle$ of the generalized hole
multiplet (\ref{vecodd})
\begin{mathletters}
\begin{eqnarray}
&& \vert \Omega_b \rangle = (\cos{\theta\over2})^{2q}
   \exp({\rm tg}{\theta\over2}e^{i\phi}{Q}_-)
   \vert B,0 \rangle\, ,  \label{cohstateven} \\
&& \vert \Omega_f \rangle = (\cos{\theta\over2})^{2q-1}
   \exp({\rm tg}{\theta\over2}e^{i\phi}{Q}_-)
   \vert F,0 \rangle\, . \label{cohstatodd}
\end{eqnarray}
\end{mathletters}
The states (2.13) are orthonormal and by generalizing standard
results \cite{rad} on su(2) coherent states we find
\begin{mathletters}
\FL
\begin{eqnarray}
&& \langle \Omega_b \vert {Q}_{\mu} \vert \Omega_b \rangle
= q(1,{\vec n}) = p_{\mu} , \;\;
\langle \Omega_b \vert {Q}_{\mu} \vert \Omega_f \rangle = 0, \label{expeven} \\
&& \langle \Omega_f \vert {Q}_{\mu} \vert \Omega_f \rangle
   = \left( q+{1\over 2},(q-{1\over 2}){\vec n} \right)
  = {p \prime}_{\mu}.  \label{expodd}
\end{eqnarray}
\end{mathletters}
As for the su(2) coherent states, the classical $c$-numbers (\ref{classvec})
are equal to the expectation value of the even generators between the even
coherent states. The spin expectation value
$ \langle \Omega_f \vert {\vec Q} \vert \Omega_f \rangle $ is in general
finite and it is zero only for the value $q={1\over 2}$, an obvious
consequence of the fact that the generalized hole is actually a
spin-$(q-{1\over2})$ multiplet.
Concerning the odd generators, it is easily checked that
\begin{eqnarray}
&& \chi_1 \vert \Omega_b \rangle =  \sqrt{2q} \, e^{i(\omega-{\phi\over2})}
   \cos({\theta\over2})\vert \Omega_f \rangle\, , \nonumber \\
&& \chi^1 \vert \Omega_f \rangle =  \sqrt{2q} \, e^{-i(\omega-{\phi\over2})}
   \cos({\theta\over2})\vert \Omega_b \rangle\, , \nonumber \\
&& \chi_2 \vert \Omega_b \rangle =  \sqrt{2q} \, e^{i(\omega+{\phi\over2})}
   \sin({\theta\over2})\vert \Omega_f \rangle\, , \nonumber \\
&& \chi^2 \vert \Omega_f \rangle =  \sqrt{2q} \, e^{-i(\omega+{\phi\over2})}
   \sin({\theta\over2})\vert \Omega_b \rangle\, , \nonumber \\
&& \chi_a \vert \Omega_f \rangle =  0\, ,  \;\;\;\;\;
   \chi^a \vert \Omega_b \rangle =  0 \label{spinlessrot}
\end{eqnarray}
where we have fixed the relative normalization of the states (2.5)
by using the phase convention $\chi_1 \vert B,0 \rangle =
\sqrt{2q} \, e^{i(\omega-{\phi\over2})} \vert F,0 \rangle$. From
Eq.\ (\ref{spinlessrot}) follows that the combination of odd generators
\begin{equation}
\chi^{\dagger} = {e^{-i\omega}\over \sqrt{2q} }\left(
                  e^{i{\phi\over2}} \cos({\theta\over2}) \chi_1 +
                  e^{-i{\phi\over2}}\sin({\theta\over2}) \chi_2 \right)
\label{spinlesscre}
\end{equation}
creates the odd coherent state out of the even one, whereas
the conjugate operator annihilates it
\begin{mathletters}
\begin{equation}
 \chi^{\dagger}\vert \Omega_b \rangle= \vert \Omega_f \rangle ,
  \;\;
 \chi          \vert \Omega_b \rangle= 0 \label{creaanni}
\end{equation}
We also have
\begin{equation}
 \chi          \vert \Omega_f \rangle= \vert \Omega_b \rangle ,
 \;\;
 \chi^{\dagger}\vert \Omega_f \rangle= 0 .
\label{parthole}
\end{equation}
\end{mathletters}
The relations (\ref{creaanni}) show that $\chi^{\dagger}$ and $\chi$ are the
$q$-generalized hole creation and annihilation operators, respectively, and
Eq.\ (\ref{parthole}) acquires the meaning of the particle/hole transformation.
Note that $\chi^{\dagger}$, $\chi$ are not spinless fermion operators,
because for a generic $q$-representation their anticommutator (restoring
the notation with the site index) is
$\{\chi^{\dagger}_i,\chi_j\} = \delta_{ij}
 {1\over2q}({Q}_{0j}+{\vec {Q}_j}{\vec n}_j)$.
However, this means that the identity
\begin{equation}
$$
 \langle{\bf \Omega}\vert \{\chi^{\dagger}_i,\chi_j\} \vert{\bf \Omega}\rangle
       =  \delta_{ij}
\label{comm}
\end{equation}
is satisfied, so that $\chi^{\dagger}_i$, $\chi_i$ act like spinless fermion
operators when bracketed between any tensor product of even and/or odd
coherent states
$\vert {\bf \Omega} \rangle = \bigotimes \vert \Omega_{\alpha i} \rangle$
($\alpha=b,f$).

\section{ LARGE-SPIN LIMIT }
\label{sec:large-spin}

Here and in the following we shall refer to the {\sl large-spin limit}
whenever the single-site Hilbert space of a strongly correlated electron
system is enlarged by assuming a spin-$q$ multiplet as the spin state and
a spin-$(q-{1\over 2})$ multiplet as the hole state, and then one lets
$q \to \infty$.
To develop the coherent states for spl(2,1) we have constantly used the
operator $Q_{0i}$, because this is the generator by which the spl(2,1)
algebraic
rules (2.4) can be closed and thus the spl(2,1) representation theory (2.5)
applied. However in the {\it t}-{\it J} model hamiltonian there enters the
charge operator ${\hat N}_i$, and we have to define its large-$q$
generalization. As a matter of fact, here we are faced with the more general
problem of how to generalize the {\it t}-{\it J} model itself, because
the large-$q$ generalization of the charge and spin operators has a certain
amount of arbitrariness. On one hand, to perform the large-spin limit of
the {\it t}-{\it J} model we may straightforwardly use the linear
generalization of the operators (2.2). This amount to replace in
Eq.\ (\ref{tjmod})
\begin{equation}
{\tilde c}_{ai} \to \chi_{ai},  \quad
{\vec S}_i \to {\vec Q}_i,      \quad
{\hat N}_i \to C_i=2(2qI-Q_{0i}),
\label{linear}
\end{equation}
where $(\chi_{ai},Q_{\mu i})$ have now arbitrary isospin magnitude $q$.
The last modification generalizes the relation $C_i=2(I-Q_{0i})$ of the
original $q={1\over2}$ theory. We have
$C_i \vert \Omega_{bi} \rangle =   2q   \vert \Omega_{bi} \rangle $, and
$C_i \vert \Omega_{fi} \rangle = (2q-1) \vert \Omega_{fi} \rangle $,
which means that the spin and hole states (2.5)
have generalized spin -- differing by half-unit --
as well as generalized electric charge -- differing by one unit.

On the other hand, for the $q={1\over2}$ fundamental representation, and
only in this case, the $3\times 3$ Hubbard matrices (\ref{hubbmat}) satisfy
\FL
\begin{eqnarray}
&& C_i = (\chi^1_i\chi_{1i}+\chi^2_i\chi_{2i}), \;
   Q_{3i} = {1\over2} (\chi^1_i\chi_{1i}-\chi^2_i\chi_{2i}), \nonumber \\
&& \label{fundam}  \\
&& Q_{1i} = {1\over2} (\chi^1_i\chi_{2i}+\chi^2_i\chi_{1i}), \;
   Q_{2i} = {1\over2i}(\chi^1_i\chi_{2i}-\chi^2_i\chi_{1i}), \nonumber
\end{eqnarray}
a consequence of the fact that whether the constraint of no doubly occupancy
is enforced, the charge and spin operators (\ref{evenop}) are unchanged if we
substitute the electron operators with the corresponding hole operators
(\ref{oddop}).
According to the previous equation, we may think of the charge and spin
operators (\ref{evenop}) as composite operators of the odd generators for any
value of $q$, so that we may also define the large-spin limit of
Eq.\ (\ref{tjmod}) by means of the replacement
\FL
\begin{equation}
{\tilde c}_{ai} \to \chi_{ai}, \;{\vec S}_i \to {\vec {\cal Q}}_i =
{1\over 2} \chi^a_i{\vec \sigma}_{ab}\chi_{bi},  \;
{\hat N}_i \to {\cal C}_i = \chi^a_i\chi_{ai},
\label{nonlinear}
\end{equation}
where $\chi_{ai}$, $\chi^a_i$ have now arbitrary isospin magnitude $q$.
The present generalization amounts to consider Eq.\ (\ref{tjmod}) as a model
of interacting holes and to preserve this dynamics for any value of the hole
isospin. For any $q\ne{1\over 2}$ Eqs.\ (\ref{linear},\ref{nonlinear}) are not
equivalent and in our analysis we shall investigate the physical implications
underlying the two choices.

The simplest generalization is obtained by choosing the linear realization
(\ref{linear}). Using the arbitrarily rotated frame $(P_{\mu},X_a)$
(\ref{eventransf},\ref{oddtransf}), upon the substitution (\ref{linear})
the generalized {\it t}-{\it J} model hamiltonian becomes
\FL
\begin{eqnarray}
  H_{tJ}^{(q)} \equiv  &&  t\sum_{ij,a}\Omega_{ij} X^a_iX_{aj} +
  {J\over 2} \sum_{ij}\Omega_{ij}\left({\vec P}_i{\vec P}_j-
  {1\over 4}C_{i}C_{j}\right)  \nonumber \\
               =  && t\sum_{ij,a}\Omega_{ij} X^a_iX_{aj} -
                     {J\over 2} \sum_{ij, \mu\nu}\Omega_{ij}
                     g^{\mu\nu}P_{\mu i}P_{\nu j} + \nonumber \\
                  && 4qJN_h, \label{tjlin}
\end{eqnarray}
where in the last identity we have applied the conservation law
(\ref{conserv}).
Similarly to spin systems, using the linear generalization we preserve all the
symmetries of the original model. Apart from a trivial constant,
Eq.\ (\ref{tjlin}) is bilinear in the spl(2,1) generators, so that $J=2t$
remains the ``supersymmetric point'' \cite{susy} for any value of the isospin
$q$.
Replacing the HP map (\ref{hpmap}) in Eq.\ (\ref{tjlin}), in the
$q \to \infty$ limit the potential energy becomes dominant, so that the
large-spin limit of Eq.\ (\ref{tjlin}) is given by the classical Heisenberg
model
\begin{equation}
H_{tJ}^{(q)} \rightarrow
H_H^{\rm class} = {Jq^2\over 2} \sum_{ij}\Omega_{ij}
\left({\vec n}_i{\vec n}_j-1\right).
\label{heisenberg}
\end{equation}
Thus for any electron density
$\rho = N_{\rm el}/N $ the mean-field solution is always the Ne\'el
spin configuration. We can easily understand this result noting that in the
large-$q$ limit the expectation values (\ref{expeven},\ref{expodd}) are equal
\begin{equation}
   \lim_{q\rightarrow\infty} {{p_{\mu i}}\over q} =
   \lim_{q\rightarrow\infty} {{{p \prime}_{\mu i}}\over q} = (1,{\vec n}_i).
\label{explimit}
\end{equation}
As a consequence in the large-$q$ limit we still have two distinct states
$\vert \Omega_{bi} \rangle$ and $\vert \Omega_{fi} \rangle$ differing by the
fermion number $F_i$, however their spin and charge become equal; hence
the mean-field theory turns in the AF spin hamiltonian,
independently from the density.
Thus in the large-spin limit we preserve all the symmetries but we miss the
following important property of the exact model (\ref{tjmod}):
if the hole is localized on site $i_0$, the four bonds
$J\Omega_{i_0 j}$ do not contribute to the energy (we shall refer to this
property as the ``dynamics of the missing bonds'').
This is a peculiar behaviour of the present large-spin limit.
However all the developments are mathematically well definite and the
expansion in fluctuations about the AF mean-field solution can be performed
straightforwardly. We divide the lattice into two sublattices and in
Eqs.\ (\ref{oddtransf},\ref{eventransf}) set
$\theta=\phi=\psi=\gamma=0$ on one sublattice, and $\theta=\psi=\pi$,
$\phi=0$, $\gamma=-\pi/2$ on the other one. Then the replacement in
Eq.\ (\ref{tjlin}) of the HP realization (\ref{hpmap}) expanded in powers
of $1/q$ gives the effective hamiltonian
\FL
\begin{eqnarray}
H^{\rm KLR} =  && E_{0}+{qJ\over2}\sum_{ij}\Omega_{ij}
                     \left(a_i^{\dagger}a_i+a_j^{\dagger}a_j+
                     a_i^{\dagger}a_j^{\dagger}+a_ia_j\right) \nonumber \\
                  && - {\sqrt{2q}}~t\sum_{ij}\Omega_{ij}
                       c_i^{\dagger}c_j(a_j^{\dagger}+a_i)+... \label{klrham}
\end{eqnarray}
to describe fluctuations, where
\begin{equation}
E_0=-4NJq^2+4Jq\sum_i c^{\dagger}_i c_i
\label{eclass}
\end{equation}
is the classical energy of the pure antiferromagnet plus a quantum
contribution which shifts the chemical potential of the spinless fermions.
In Eq.\ (\ref{klrham}) we have displayed only the first few terms of the
systematic expansion. They give exactly the hamiltonian proposed and applied
by Kane, Lee, and Read (KLR) \cite{klr}
to describe the propagation of a single hole in a quantum antiferromagnet.
In their approach the bosonic term and the three-body term -- the latter
responsible for the hole propagation through spin-wave emission and
absorbtion -- are treated on equal footing. This assumption corresponds to
scaling $t$ by a factor ${\sqrt{2q}}$ and then define the large-spin limit
keeping fixed the renormalized hopping parameter $t_0=t/{\sqrt{2q}}$.
Eq.\ (\ref{klrham}) is well definite for any density and from
Eq.\ (\ref{eclass}) we see that in the large-spin limit (\ref{linear})
the dynamics of the missing bonds is present
only when quantum fluctuations are taken into account. In particular setting
$q={1\over2}$, we have $E_0=-J(2N-4N_h)/2$, which is the energy of the
classical antiferromagnet with $N_h$ delocalized holes.

The dynamics of the missing bonds is the essential feature of the
{\it t}-{\it J} model to understand phase separation in cuprate
superconductors, and it is completely not accounted for by the large-spin
limit (\ref{linear}). This property can be preserved if we resort to the
nonlinear generalization (\ref{nonlinear}). All our considerations are
simplified by introducing the projector onto the even sector
\FL
\begin{equation}
\Gamma_i = (2q+1)I-2Q_{0i}, \quad
\Gamma_i  \vert \Omega_{bi} \rangle = \vert \Omega_{bi} \rangle , \quad
\Gamma_i  \vert \Omega_{fi} \rangle = 0. \label{project}
\end{equation}
Employing the HP transformation (\ref{hpmap}), it is easily shown that the even
operators $({\cal C}_i,{\cal {\vec Q}}_i)$ of the set (\ref{nonlinear})
are related to the operators $({ C}_i,{ {\vec Q}}_i)$ in Eq.\ (\ref{linear})
via the projector $\Gamma_i$
\begin{equation}
{\cal C}_i = \Gamma_i^{\dagger} C_i \Gamma_i , \quad\quad
{\vec {\cal Q}}_i = \Gamma_i^{\dagger} {\vec Q}_i\Gamma_i .
\label{linnonlin}
\end{equation}
We collect the generalized charge and spin operators (\ref{nonlinear})
in the four-vector ${\cal Q}_{\mu i} = \Gamma_i^{\dagger} Q_{\mu i} \Gamma_i$
(note that we have the relation
${\cal C}_i =\Gamma_i^{\dagger} C_i \Gamma_i = 2 {\cal Q}_{0i} $).
Applying Eqs.\ (\ref{project},\ref{linnonlin}) we then have
\begin{equation}
\langle\Omega_{bi}\vert {\cal Q}_{\mu i} \vert\Omega_{bi}\rangle = p_{\mu i},
\quad\quad
\langle\Omega_{fi}\vert {\cal Q}_{\mu i} \vert\Omega_{fi}\rangle = 0 .
\label{expnew}
\end{equation}
An independent check of Eq.\ (\ref{expnew}) can be obtained by using the
definition (\ref{nonlinear}) and the rules (\ref{spinlessrot}). Since the
projector (\ref{project}) commutes with the even generators $Q_{\mu i}$
it is easily proved that the su(2)$\times$u(1) even subalgebra is preserved
\begin{mathletters}
\begin{equation}
  [ {\cal Q}_{\mu i}, {\cal Q}_{\nu j}] =
  i\delta_{ij}{\epsilon_{0\mu\nu}}^{\lambda} {\cal Q}_{\lambda j},
\label{subalgebra}
\end{equation}
so that the operators (\ref{linnonlin}) under the rotation (\ref{oddtransf})
of the odd generators, $\chi_{ai} \to X_{ai}$, are
$({\cal P}_{\mu i},X_{ai})$, where
\begin{equation}
  {\cal P}_{0i}={\cal Q}_{0i}={1\over 2}{\cal C}_{0i}, \qquad
  {\cal P}_{ki}= X_i^a{\sigma}_{ab}^k X_{bi}=R_{km}{\cal Q}_{mi},
\label{rotframe}
\end{equation}
\end{mathletters}
$R_{km}$ being the rotation matrix in Eq.\ (\ref{eventransf}). Upon the
substitution (\ref{nonlinear}) the generalized {\it t}-{\it J} model
hamiltonian in the rotated frame (\ref{rotframe}) eventually becomes
\begin{eqnarray}
H_{[X,{\cal P}]}^{(q)}
  = && t\sum_{ij,a}\Omega_{ij}X_i^aX_{aj}+ \nonumber \\
    && {1\over2}{J_0\over2q}\sum_{ij}\Omega_{ij}
       \left({\vec {\cal P}}_i{\vec {\cal P}}_j-
       {1\over 4} {\cal C}_{i}{\cal C}_{j}\right) .
\label{tjnonlin}
\end{eqnarray}
Here we have defined the model keeping fixed the rescaled exchange constant
$J_0 = 2q J$, to weight on equal footing the kinetic and potential energy
contributions (we explicitly display the dependence of
$H_{[X,{\cal P}]}^{(q)}$ on the operators for later convenience).

For arbitrary values of $q$ we loose supersymmetry at the $J=2t$ point
because the set of operators $({\cal Q}_{\mu i},\chi_{ai})$ forms a
representation of spl(2,1) only for $q={1\over 2}$. However due to Eqs.~(3.12)
the model (\ref{tjnonlin}) remains rotationally invariant and, thanks
to Eq.\ (\ref{expnew}), the dynamics of the missing bonds is now present for
any value of the isospin $q$. Thus Eq.\ (\ref{tjnonlin}) has all the relevant
properties of the {\it t}-{\it J} model we are interested in and we consider
it as a physically sensible generalization.

In the canonical basis the operators (\ref{linnonlin}) are given by the
following expressions
\begin{eqnarray}
&&  {\cal C}_i = 2qc_ic^{\dagger}_i , \;\;
    \chi_{1i} = c^{\dagger}_i\sqrt{2q-a^{\dagger}_ia_i},  \nonumber \\
&&  {\cal Q}_{3i} = c_ic^{\dagger}_i(q-a^{\dagger}_ia_i) , \;\;
    \chi^1_i = \sqrt{2q-a^{\dagger}_ia_i}~c_i, \nonumber \\
&&  {\cal Q}_{+i} = c_ic^{\dagger}_i\sqrt{2q-a^{\dagger}_ia_i}~a_i, \;\;
    \chi_{2i} = c^{\dagger}_ia_i, \nonumber \\
&&  {\cal Q}_{-i} =
c_ic^{\dagger}_ia^{\dagger}_i\sqrt{2q-a^{\dagger}_ia_i},\;\;
    \chi^2_i = a^{\dagger}_ic_i . \label{hpnonlin}
\end{eqnarray}
Expanding in powers of $1/q$ the previous realizations, the large-spin limit
of Eq.\ (\ref{tjnonlin}) turns in an interacting
spinless fermion hamiltonian imbedded in a classical spin background
\FL
\begin{eqnarray}
H_{[X,{\cal P}]}^{(q)} \to
      &&  H^{\rm eff}_{tJ} = -2qt\sum_{ij}\Omega_{ij}
           \langle \Omega_{bi}\vert\Omega_{bj}\rangle^{1\over{2q}}
           c_i^{\dagger}c_j + \nonumber \\
      &&   {{qJ_0}\over 4}\sum_{ij}\Omega_{ij}
           \left({\vec n}_i{\vec n}_j-1\right)
           (1-c_i^{\dagger}c_i)(1-c_j^{\dagger}c_j) , \nonumber \\
\label{spinlessham}
\end{eqnarray}
where the overlap between even coherent states is
\FL
\begin{equation}
\langle \Omega_{bi}\vert\Omega_{bj}\rangle =
\left( \cos{\theta_i\over2}\cos{\theta_j\over2} +
e^{-i(\phi_i-\phi_j)}\sin{\theta_i\over2}\sin{\theta_j\over2} \right)^{2q},
\label{overlap}
\end{equation}
(so that the factor
$\langle \Omega_{bi}\vert\Omega_{bj}\rangle^{1/(2q)}$ in
Eq.\ (\ref{spinlessham}) is $q$-independent), and we have chosen the relative
normalization of the multiplets (2.5) such that $\omega_i-\phi_i/2=0$.

A relevant feature of the effective hamiltonian Eq.\ (\ref{spinlessham})
is that it gives a variational estimate of the ground-state energy. Any
further inclusion of $1/q$-fluctuations can only improve the estimate,
and this is a remarkable property that in our approach is naturally
preserved. To show this property, we consider the model (\ref{tjnonlin}) in
the original frame, namely with the operators
$({\cal P}_{\mu},X_a)$ directly replaced by $({\cal Q}_{\mu},\chi_a)$,
and denote it with $H_{[\chi,{\cal Q}]}^{(q)}$. According to Eqs.\
(\ref{spinlessrot},\ref{spinlesscre}) and to the generalization
(\ref{nonlinear}) of the charge and spin operators (\ref{evenop}) the action
of the hamiltonian $H_{[\chi,{\cal Q}]}^{(q)}$ over the state
\begin{equation}
\vert \Psi^{(q)} \rangle =
\prod \limits_{l=1}^{N_h} \left( \sum_{i=1}^N\psi_l (i) \chi^{\dag}_i \right)
\vert {\bf \Omega} \rangle  ,
\label{state}
\end{equation}
is exactly given by the action of the effective spinless fermion hamiltonian
(\ref{spinlessham}) over the free particle state $\vert \Psi \rangle$
obtained by replacing in Eq.\ (\ref{state}) the operators $\chi^{\dagger}_i$
with spinless fermion operators $c^{\dagger}_i$ and the
boson state $\vert {\bf \Omega} \rangle$ with the vacuum state
$\vert 0 \rangle $ of the spinless fermion representation. Here
$\vert {\bf \Omega} \rangle = \bigotimes \vert {\Omega}_{bi} \rangle$ is
the tensor product of even coherent states and $\psi_l(i)$ are $N_h$ complex
and orthonormal orbitals ($l=1,2,...,N_h$ in a $d$-dimensional hypercubic
lattice $L^d=N$).

As a consequence, any variational estimate
${\cal E} = \langle \Psi \vert H^{\rm eff}_{tJ} \vert \Psi \rangle $
of the ground-state energy of the spinless fermion hamiltonian
(\ref{spinlessham}) is also a variational estimate of the ground-state energy
of the generalized {\it t}-{\it J} model (\ref{tjnonlin}). Moreover simple
algebra shows the notable identity
\begin{equation}
{\cal E} =
\langle \Psi^{(q)} \vert H^{(q)}_{[\chi,{\cal Q}]} \vert \Psi^{(q)} \rangle =
\, 2q \,
\langle \Psi^{({1\over2})} \vert H_{tJ} \vert \Psi^{({1\over2})} \rangle ,
\label{variational}
\end{equation}
where $H_{tJ}$ is the exact $q={1\over 2}$ model (\ref{tjmod}). The factor
$2q$ simply rescales the unit of energy and this ensures that no spurious phase
transitions as a function of $q$ are present in connecting the large-spin
``classical'' solution ${\cal E}$ to the variational estimate of the exact
model. For the time being we thus set $2q=1$.

\section{ LARGE-SPIN MEAN-FIELD  }
\label{sec:mean-field}

Despite the notable simplification given by the disappearance of the
constraint of double occupancy, Eq.\ (\ref{spinlessham})
is still a difficult problem to solve because the spinless fermions
interact via a four-body term. For the infinite-$U$ Hubbard model, i.e.,
when $J=0$, and for the case of a single hole the interaction term does not
play any role. Thus in the large-$q$ limit we end up with the study of a
free hamiltonian with a complicated site dependent hopping, which
can be easily solved numerically on a finite -- but large -- lattice, as we
shall discuss. For low-doping the Hartree-Fock factorization (\ref{state})
is expected to be a good approximation since few holes can rarely interact.
We extended the Hartree-Fock solution to the full phase diagram of the
hamiltonian (\ref{spinlessham}). Although this is not completely justified
the interaction between spinless fermions should not drastically affect the
mean-field phase diagram,
since for example it is known that the the Hartree-Fock solution is exact in
$d\to \infty$ for a gas of interacting spinless fermions \cite{voll}.

Thanks to Eq.\ (\ref{variational}) it is possible to apply Wick's theorem and
the classical energy on a given arbitrary state reads
\FL
\begin{eqnarray}
{\cal E} = &&  \sum_{ij}  \Omega_{ij}  \bigl[
   - t \langle \Omega_{bi}\vert\Omega_{bj}\rangle g_{i,j}\bigr. \nonumber \\
&& - \bigl. {J\over 8 } ({\vec n}_i {\vec n}_j -1)(  g_{i,i} + g_{j,j}-
g_{i,i} g_{j,j} + g_{i,j} g_{j,i}-1)  \bigr]; \nonumber \\
&& \label{functional}
\end{eqnarray}
where $g_{i,j}=\sum_{l=1}^{N_h} \psi^*_l(i) \psi_l(j)$.

The classical energy is therefore a function of the $2N$ spin angles
$\theta_i$, $\phi_i$ and of the $N_h\times N$ complex variables defining
the wavefunction (\ref{state}). The lowest possible energy ${\cal E}_0$ as
a function of all the $2N\times (1+N_h)$ real
variables represents a variational classical estimate of the ground-state
energy. We solved this optimization problem on a square lattice numerically,
following a scheme which is similar to the one introduced by Car and
Parrinello \cite{car} for the electronic structure problem, i.e., for the
simulation of the {\sl slow} dynamics of the ions interacting via a
self-consistent potential generated by the electronic {\sl fast}
degrees of freedom. This approach can be extended to our case because
the isospin $q$ behaves as an adiabatic parameter: in the
$q\to\infty$ limit the spl(2,1) even generators (\ref{hpmap})
(``slow variables'') become classical objects, so that Eq.\ (\ref{tjnonlin})
describes the dynamics of the odd and projection operators (\ref{hpmap}),
(\ref{project}) (``fast variables'') in the background of the former.
Thus the electronic degrees of freedom have a much faster dynamic of the spin
angles, which in this case play indeed the role of the ionic coordinates.

In order to minimize the numerical effort we first move the electronic
degrees of freedom at a fixed spin configuration, and then the spin angles
without changing the electronic degrees of freedom. At each step we do not
require a fully converged Hartree-Fock solution of the electronic part, but
we make a fixed number of steepest descent steps, followed by a
Graham-Schmidt orthogonalization of the orbitals to achieve numerical
stability. After many spin-moves ($ \approx 10,000$)
we get a fully self consistent solution of the problem up to computer
machine accuracy. The solution found with the previously described
iterative scheme may not coincide with the absolute minimum -- the classical
ground-state. The true minimum can be identified on a reasonable ground by
resorting to symmetry considerations or by performing several simulations
with different random initializations.

For $J=0$ the mean-field hamiltonian (\ref{spinlessham}) becomes the free
spinless fermion hopping hamiltonian
\begin{equation}
H^{\rm eff }_t = -t\sum_{ij}\Omega_{ij}
                   \langle \Omega_{bi}\vert\Omega_{bj}\rangle
                   c_i^{\dagger}c_j,
\label{wenham}
\end{equation}
which has been phenomenologically introduced \cite{wen} to
study the instability of the Nagaoka state -- i.e., the free particle state
with maximum allowed spin and lowest kinetic energy -- in the infinite-$U$
Hubbard model. The Nagaoka theorem \cite{nagaoka} states that it is
the exact ground-state of the model in the one hole case. Writing the
coherent state overlap in the form \cite{rad}
\begin{equation}
  \langle \Omega_{bi} \vert \Omega_{bj} \rangle =
  \sqrt{ {{1+{\vec n}_i{\vec n}_j}\over 2} } \, e^{i A_{ij} }
\label{flux}
\end{equation}
where $A_{ij}$ is the solid angle subtended by the vectors ${\vec n}_i$,
${\vec n}_j$, and a reference fixed one ${\vec n}_0$, it has been conjectured
that for more than one hole a non-fully polarized spin background may have
better energy than the Nagaoka state, because the gain in magnetic energy
due to the nonzero flux of $A_{ij}$ could overwhelm the narrowing of the
effective bandwidth
$t\Omega_{ij} \to t\Omega_{ij} \sqrt{ ({1+{\vec n}_i{\vec n}_j})/ 2 }$.
In this case we have performed over $10,000$ fully converged
minimizations on a $10 \times 10$ and a $16 \times 16$ lattice and for
arbitrary density. We initialize randomly the spins angles  and
set the orbitals $\psi_l(i)$ at the corresponding Hartree-Fock solution of the
spinless-fermion hamiltonian with the chosen random spin configuration.
In all the minimizations we have {\sl never} found a solution with
corresponding energy lower than the ``Nagaoka energy''.
Most of the runs converge to the fully polarized solution. Only a few remain
trapped into a local minimum of the classical energy.
A particularly interesting case is at doping $\delta=N_h/N={1/2}$.
In this case we have indeed found a stable planar solution with $\phi_{i}=0$
and $\theta_{i}={2\pi \over L}i_x$, where $i_x\,=\, 0,1,...,L-1$ are the
lattice coordinates along the $x$-axis. This minimum configuration is similar
to the one proposed by Douc\c{o}t and Wen for few holes \cite{wen}.
However we have found that this kind of state is always unstable except
for this particular doping $\delta ={1/2}$, and its energy is only degenerate
with the Nagaoka energy.
For large $L$ this state tends (locally) to the Nagaoka state and leads
in fact to the same correlation functions. Contrary to the Nagaoka
state, this planar solution is not an exact eigenstate of the
infinite-$U$ Hubbard model. This fact represents a very simple proof that the
Nagaoka state is not the true ground-state at $\delta ={1/2}$.
Although our result about the stability of the Nagaoka state is based
on a numerical optimization problem which is never completely reliable,
we at least may argue that there is no evidence of any exotic
spiral, flux or chiral phase in the {\it t}-{\it J} model for $J=0$
at the mean-field level.

For $J\ne 0$ we consider first the interesting case when only one hole is
present. A recent reason of debate is whether the spiral mean-field solution
of Shraiman and Siggia \cite{ss} has lower or higher energy compared to the
polaronic solutions (see Fig.~\ref{fig1}). As we have anticipated, also in this
case we are left with the study of a free hamiltonian with a complicated
site-dependent hopping, to be adjusted as to minimize Eq.\ (\ref{functional}).
The results of our numerical
investigation confirm that the polarons are the lowest-energy configurations
of the large-spin limit of the {\it t}-{\it J} model. The small-polaron
solution, also called five-sites polaron (the hole is trapped in a given site
and its nearest neighbors: see Fig.~\ref{fig1}a), is stable for $J/t>0.243$.
For smaller $J/t$ the size $\xi$ of the polaron is gradually increased (the
8-sites polaron in Fig.~\ref{fig1}b is stable for $0.148<J/t<0.243)$ until
$\xi \sim J^{-{1\over 2}}$, consistent with theoretical arguments
\cite{emery}, so that in the $J/t \to 0$ the polaron solution
eventually turns in the Nagaoka state.
With the polaron spin background the mean-field energy and the orbital can be
easily evaluated analytically. We report the case of the five-sites polaron
localized at site $i_0$. Writing the orbital in Eq.\ (\ref{state}) in the form
$\psi(i)=f\delta_{i_0 i}+g_i\Omega_{i_0 i}$, the diagonalization of
Eq.\ (\ref{spinlessham}) gives
\begin{eqnarray}
&& {\cal E}_0 = -{J\over2}(2N-4)-2t\lambda_0, \nonumber \\
&& \lambda_0= -{3J\over 8t}+
   {\rm sign}(t)\,{\sqrt{1+\left({3J\over 8t}\right)^2 }} \nonumber \\
&& f = {1\over {\sqrt{1+\lambda_0^2}} }, \qquad\qquad
   g_i = {\lambda_0\over {2{\sqrt{1+\lambda_0^2}}} }
\label{polaron}
\end{eqnarray}
The energy of the antiferromagnet with four missing bonds is recovered
in the $J/t \to \infty$ limit, because we have $\lambda_0 \to 0$ so that
$f \to 1$, $g_i \to 0$, which means that the hole is statically placed on
site $i_0$.
The one-hole energy (referred to the half-filled ground-state energy
${\cal E}_{\rm AF}=-J N$) for the Shraiman and Siggia state at momentum
${\bf k}=({\pi \over 2},{\pi \over 2})$ and the polaron estimate is
plotted in Fig.~\ref{fig2} as a function of $J/t$ in a $10 \times 10$ lattice.
The comparison of the two energies is meaningful because both represent
a variational estimate of the ground-state energy of the {\it t}-{\it J}
model. However the two variational wavefunctions refer to different
order of approximation, because the hole in the Shraiman and Siggia
estimate has a definite momentum.
Propagating the polaron through the whole lattice via processes as
depicted in Fig.~\ref{fig3} (highly non-perturbative in $1/q$, as discussed in
Ref. \cite{aul2}), eventually decreases the energy, but the present data
shown in the Fig.~\ref{fig2} clearly indicate that the spiral has lower energy
than the static polaron, unless for small $J/t$.

We consider next the case of four holes. For large $J/t$ the stable state is
clearly localized in a $2 \times 2$ plaquette (see Fig.~\ref{fig1}c). As we
decrease $J$, for $J/t< 1.09 $ the four holes prefer to
remain localized in a polaron of $13$ sites, tilted by $45$ degrees respect to
the $x,y$ axes (see Fig.~\ref{fig1}d), rather than to split into a couple of
bound pairs. For $J/t$ smaller and smaller
the size of the polaron smoothly increases until the spins are fully polarized
in the given finite $10\times10$ square lattice.
These results  support the Emery {\it et al.} scenario. At least at the mean
field level  the classical solution consists of an hole-rich phase fully
polarized and a classical antiferromagnetic region which are completely
phase separated for arbitrary $J$. The exact phase boundary in the
thermodynamic
limit may be obtained by minimizing the energy of the phase separated phase.
Following Emery {\it et al.} we get a critical $J_c$ above which phase
separation occurs, given by
\begin{equation}
{ {J_c(\delta)}\over {\vert t \vert} }
\, =\, { \int \limits_{-4\vert t \vert}^{E_F} (E_F-E) N(E) dE \over 2B},
\label{jcritic}
\end{equation}
\FL
\begin{equation}
\delta = \int \limits_{-4\vert t \vert}^{E_F}  N(E) dE ,
\;
N(E)={1\over {2 \pi^2 \vert t \vert} } K(\root \of {1-(E/4t)^2});
\label{ellit}
\end{equation}
where the spinless fermion density of states $N(E)$ is expressed
in terms of  the complete elliptic integral of the first kind $K$, $E_F$ is
the spinless fermion Fermi energy at the corresponding doping $\delta$, and
$B$ is the classical energy per bond ($B=1$) of the Heisenberg
antiferromagnet. The resulting phase diagram in the $J-\rho$ plane, where
$\rho=1-\delta$, is shown in Fig.~\ref{fig4}. Here we note the characteristic
$\delta \sim \sqrt {J}$ singularity occurring at low doping as the size
$\xi \sim J^{-{1\over 2}}$ of the polaron increases in a singular way.
At low density Eq.\ (\ref{jcritic}) is valid only if electrons do not
form bound states. This is actually the case for the exact {\it t}-{\it J}
hamiltonian \cite{emery}, where two electrons form a bound singlet pair for
$J/t > 2$. In this approach however the paramagnetic phase
at low density is clearly not well characterized  since the spins are
frozen in some fixed  direction. As a result two electrons never bind
-- as we have tested numerically -- and the Emery-like phase boundary
represents the exact mean-field phase diagram.
Let us discuss now what happen for $J\le J_c$ at fixed density.
At $J=J_c$ the Nagaoka state
is stable by construction since the hole-rich phase exhausts all the allowed
space at the given density. It is then easy to convince ourselves that if
we decrease $J/t$ the Nagaoka state is even more stable because lowering
$J/t$ favors the polarized solution.

At the large-spin {\sl mean-field level} we therefore end up with a very
simple phase diagram consisting of a ferromagnetic phase for $J \le J_c$
and a separated phase for $J > J_c$.

\section{ DISCUSSION AND CONCLUSIONS }
\label{sec:discuss}

In this paper we have presented the coherent states that allow to give
a variational foundation to the large-spin limit of the {\it t}-{\it J}
model. These states are obtained by grading according to the spl(2,1)
algebra the standard su(2) spin coherent states. We have emphasized how
the large-spin limit can be performed in distinct schemes, and applying one
of them we have rigorously derived the Kane, Lee, and Read hamiltonian.
We have discussed the shortcoming of this approach, related to the fact that
in this scheme the mean-field solution turns out to be the classical Ne\'el
state, independently from the density. As a consequence phase separation
cannot be recovered at the mean-field level.

We then have presented and numerically investigated the effective spinless
fermion hamiltonian obtained by means of a different and more satisfactory
definition of the large-spin generalized {\it t}-{\it J} model.
This mean-field model hamiltonian has the essential properties and symmetries
of the original hamiltonian, in particular it exactly takes into account
the constraint of no doubly occupancy.
The mathematical tools we have employed are similar to other proposed
techniques, notably the Swinger bosons/slave-fermion representation.
However we hope to have better clarified the mathematical and physical
assumptions underlying the large-spin limit of the {\it t}-{\it J} model.
More important, by means of the spl(2,1) graded
coherent states we have shown that the mean-field solution is independent
of the magnitude of the expansion parameter and therefore the corresponding
energy is variational. In this way we have generalized a very useful property
of the spin systems.

Our numerical investigation confirms that the polaron solutions are
the lowest-energy configurations of the {\it t}-{\it J} model in the
large-spin limit. However we do not find that the static small-polaron
has lower energy compared to the Shraiman and Siggia spiral state as
reported by Auerbach and Larson \cite{aul2}, at least when in the definition
of the {\it t}-{\it J} model three-site contributions are neglected.
The investigation of the full phase diagram at the mean-field level support
the Emery {\it et al.} picture: in the $J-\delta$ plane a critical
$J_c=J_c(\delta)$ with $J_c\to 0$ for $\delta \to 0$ separates the
ferromagnetic and the phase-separated phases.

For $J=0$ the large-spin solution is always fully polarized
and our numerical analysis suggests that the instability of the Nagaoka
state cannot be found at the mean-field level.
The mean-field picture is exact for a single hole, likely for small doping,
but is incorrect for large doping, where existing numerical works
\cite{luchini} and the variational singlet Gutzwiller projected wavefunction
\cite{shiba} result seems to suggest
that the Nagaoka state never survives at finite concentration of holes.
On the other hand for $J=0$ the energy ${\cal E}_0$ obtained by minimizing
the functional (\ref{functional}) has the particle-hole symmetry, i.e.,
${\cal E}_0(\delta)={\cal E}_0(1-\delta)$.
This is not a true symmetry of the infinite-$U$ Hubbard model and
thus we expect that $1/q$-fluctuations play a relevant role, especially
for small density.
Non-perturbative processes may be of some relevance, however
we expect the mean-field picture to be reliable in the large $J/t$
region.

\acknowledgements

A.A. gratefully acknowledges the Istituto Nazionale di Fisica della
Materia for financial support and Dr. R. Link for useful
discussions. Both the authors are grateful to Prof. Erio Tosatti for
many elucidating discussions. This research was supported by the Italian
Ministry of University and Research, the Istituto Nazionale di Fisica della
Materia, and the Consiglio Nazionale delle Ricerche under Progetto Finalizzato
``Sistemi Informatici e Calcolo Parallelo''.

\figure{
         Polaronic solutions. The symbols $\circ$ and ${\bullet}$
         indicate spin up and spin down, respectively. The symbols
         $\otimes$ and $\odot$ indicate spin up and spin down, and in
         addition a nonzero hole density on the site. For the one hole
         case we report the small-polaron (a), the 8-sites polaron (b),
         and the 13-sites polaron (d).
         For the four holes case it is reported
         the 2$\times$2 plaquette (c) and the 13-sites polaron (d).
         \label{fig1}  }
\figure{
         Variational hole energy in unit of $\vert t \vert$ as a
         function of $J$ on a 10$\times$10 lattice. The dashed line is
         the semiclassical
         spiral solution of Shraiman and Siggia with momentum
         ${\bf k}=({\pi\over 2},{\pi\over 2})$. The continuous line
         represents our large-$q$ mean-field solution: a static polaron
         of increasing size as $J$ is lowered. The abrupt changes
         of the energy at small $J$ are due to corresponding
         first order transitions to larger and larger polaron size.
         \label{fig2}  }
\figure{
         Second order process for the $q={1\over 2}$ theory that allows
         the small-polaron propagation. The wavefunction
         $\vert \Psi_{i} \rangle$ of the polaron at site $i$ is written
         as $\vert \Psi_{i} \rangle =
         \vert f_{i} \rangle +\sum_k \Omega_{ik} \vert g_{ik} \rangle$.
         In (a) we have the state  $\vert g_{ik} \rangle$ entering the
         polaron at site $i$. In (b) we have the intermediate state
         $ \vert \delta_{i}^{j} \rangle \sim H_{tJ}\vert g_{ik}\rangle $
         which is a one-particle excited state and corresponds to a
         local minimum of the energy functional (4.1). In (c) we have
         the state $\vert f_{j} \rangle$ entering the polaron at site
         $j$, which is obtained by a further application of the
         hamiltonian:
         $\vert f_{j} \rangle \sim H_{tJ} \vert \delta_{i}^{j} \rangle$.
         Due to
         $\langle\Psi_{i}\vert H_{tJ}^2\vert\Psi_{j}\rangle \sim g_{i}ftJ$,
         for $i \ne j$, an effective intrasublattice hopping is
         generated.
         \label{fig3}  }
\figure{
         Phase diagram of the {\it t}-{\it J} model obtained with the
         Hartree-Fock large-spin approach.
         \label{fig4}  }

\begin{references}
\bibitem{klr} C.L. Kane, P.A. Lee, and N. Read, Phys. Rev. B {\bf 39},
              6880, (1989).
\bibitem{ss} B.I. Shraiman and E.D. Siggia, Phys. Rev. Lett. {\bf 61},
             467 (1988); Phys. Rev. B {\bf 42}, 2485 (1990).
\bibitem{voll} W. Metzner, P. Schmit, and D. Vollhardt, Phys. Rev B {\bf 45},
              2237 (1992); R. Strack and D. Vollhardt, preprint RWTH/ITP-C2/92.
\bibitem{grilli} M. Grilli and G. Kotliar, Phys. Rev. Lett. {\bf 64}, 1170
                 (1990); I.R. Pimentel and R. Orbach, Phys. Rev. B {\bf 46},
                 2920 (1992).
\bibitem{numw} K.J. von Szczepanski, P. Horsh, W. Stephan, and
                M. Ziegler, Phys. Rev. B {\bf 41}, 2017 (1990);
                E. Dagotto, R. Joynt, A. Moreo, S. Bacci, and E. Gagliano,
                {\em ibid.} {\bf 41}, 9049 (1990); H. Yokoyama and H. Shiba,
                J. Phys. Soc. of Japan {\bf 56}, 3570 (1987).
\bibitem{emery} V.J. Emery, S.A. Kivelson, and H.Q. Lin, Phys. Rev. Lett.
                {\bf 64}, 475 (1990).
\bibitem{castro} M. Grilli, P. Raimondi, C. Castellani, and C. Di Castro,
                 Phys. Rev. Lett. {\bf 67}, 259 (1991).
\bibitem{ogata} M. Ogata, M.U. Luchini, S. Sorella, and F.F. Assaad,
                Phys. Rev. Lett. {\bf 66}, 2388 (1991).
\bibitem{putikka} W.O. Putikka, M.U. Luchini, and T.M. Rice,
                  Phys. Rev. Lett. {\bf 68}, 538 (1992).
\bibitem{al} A. Angelucci and R. Link, Phys. Rev. B {\bf 46}, 3089 (1992);
             in {\it Proceedings of the Conference: Field Theory and Collective
             Phenomena}, Perugia, Italy, 1992 (World Scientific,
             Singapore, to be published).
\bibitem{wen} B. Douc\c{o}t and X.G. Wen, Phys. Rev. B {\bf 40}, 2719 (1989);
              B. Douc\c{o}t and R. Rammal, {\em ibid.} {\bf 41}, 9617 (1990).
\bibitem{tjm} For a review see, e.g. : A.P. Balachandran, E. Ercolessi, G.
              Morandi, and A.M. Srivastava, Int. J. Mod. Phys. B {\bf 4}, 2054,
             (1990); {\it The Hubbard Model, A Reprint Volume}, edited by
             A. Montorsi (World Scientific, Singapore, 1992).
\bibitem{wieg} P.B. Wiegman, Phys. Rev. Lett. {\bf 60}, 821, (1988).
\bibitem{snr} M. Scheunert, W. Nahm, and V. Rittenberg, J. Math. Phys.
              {\bf 18}, 155, (1977).
\bibitem{hp} T. Holstein and H. Primakoff, Phys. Rev. {\bf 59}, 1098, (1940).
\bibitem{aul} A. Auerbach and B.E. Larson, Phys. Rev. B {\bf 43}, 7800 (1991).
\bibitem{rad} J.M. Radcliffe, J. Phys. A: Gen. Phys. {\bf 4}, 313, (1971).
\bibitem{susy} P.-A. Bares, G. Blatter, and M. Ogata, Phys. Rev. B {\bf 44},
               130 (1991).
\bibitem{car} R. Car and M. Parrinello, Phys. Rev. Lett. {\bf 55}, 2471 (1985);
             {\it Simple Molecular Systems at Very High Density}, edited by
             Polian, P. Lowbeyre, and N. Boccara (Plenum Publishing
Corporation,
             1989).
\bibitem{nagaoka} Y. Nagaoka, Phys. Rev {\bf 147}, 392 (1966).
\bibitem{aul2} A. Auerbach and B.E. Larson, Phys. Rev. Lett. {\bf 66},
               2269, (1991).
\bibitem{luchini} W.O. Putikka, M.U. Luchini, and M. Ogata, Phys. Rev. Lett.
                  {\bf 69}, 2288 (1992).
\bibitem{shiba} H. Yokoyama and H. Shiba, in Ref. \cite{numw}.
\end{references}
\end{document}